# HST OBSERVATIONS OF GIANT ARCS: HIGH RESOLUTION IMAGING OF DISTANT FIELD GALAXIES*

Ian Smail,[1] Alan Dressler,[1] Jean-Paul Kneib,[2] Richard S. Ellis,[2]
Warrick J. Couch,[3] Ray M. Sharples[4] & Augustus Oemler Jr.[5]

1) The Observatories of the Carnegie Institution of Washington, 813 Santa Barbara St., Pasadena, CA 91101-1292
2) Institute of Astronomy, Madingley Rd, Cambridge CB3 0HA, UK
3) School of Physics, University of NSW, Sydney 2052, NSW Australia
4) Dept. of Physics, University of Durham, South Rd, Durham DH1 3LE, UK
5) Yale Observatory, New Haven, CT 06511

## Abstract

We present HST imaging of eight spectroscopically-confirmed giant arcs, pairs and arclets. These objects have all been extensively studied from the ground and we demonstrate the unique advantages of HST imaging in the study of such features by a critical comparison of our data with the previous observations. In particular we present new estimates of the core radii of two clusters (Cl0024+16, A370) determined from lensed features which are identifiable in our HST images. Although our HST observations include both pre- and post-refurbishment images, the depth of the exposures guarantees that the majority of the arcs are detected with diffraction-limited resolution. A number of the objects in our sample are multiply-imaged and we illustrate the ease of identification of such features when working at high resolution. We discuss the morphological and scale information on these distant field galaxies in the light of HST studies of lower redshift samples. We conclude that the dominant population of star-forming galaxies at $z \sim 1$ is a factor of 1.5–2 times smaller than the similar group in the local field. This implies either a considerable evolution in the sizes of star-forming galaxies within the last $\sim$10 Gyrs or a shift in the relative space densities of massive and dwarf star-forming systems over the same timescale.

**Key words:** cosmology: observations – galaxies: formation – galaxies: photometry – gravitational lensing.

## 1 Introduction

The study of galaxy evolution using luminosities, sizes or other morphological characteristics of the galaxies is a standard technique in observational cosmology. Until recently, however, the limitations of observing from the ground have meant that the crucial morphological information has been unavailable for the most distant objects. This has been one of the major restrictions on the study of "normal" galaxies at high redshift. With the refurbishment of Hubble Space Telescope (HST) we are finally in a position to remove this obstacle. While HST has been used to study large samples of *cluster* galaxies at moderate redshifts (Dressler *et al.* 1994, Couch *et al.* 1994) similar sized samples of moderate and high redshift field galaxies with spectroscopic identifications have been slower to appear.

Interest in faint galaxy evolution has centred on the problem of the "excess" faint galaxy population apparent in deep optical counts (Ellis 1994). While the cumulative surface density of galaxies exceeds the extrapolations from local populations (the no-evolution model), their redshift distribution is apparently consistent, at least to $B \sim 24$, $I \sim 22$ with the same no-evolution model (Glazebrook *et al.* 1995a, Lilly 1993, Tresse *et al.* 1993). Gravitational lensing studies of the median redshift of yet fainter samples, $B \sim 27$, indicate that the no-evolution model is still an adequate description at this depth (Kneib *et al.* 1994, 1995b; Smail *et al.* 1994).

The key to understanding these results may come from morphological studies of the faint field population. A first attempt at this was made by Giraud (1992) who used ground-based imaging taken in sub-arcsecond seeing to study the morphologies of a small sample of blue $I < 23$ field galaxies. He broke his sample into three broad categories: compact galaxies with extended halos (33% of the sample), extended irregular sources (50%) and multiple or merging objects (17%). He concluded that the star-formation responsible for the blue colours of the faint field population arose from a variety of different physical mechanisms, rather than a single class of objects. A similarly simple scheme was used by Glazebrook *et al.* (1995b) and Driver *et al.* (1995) in their morphologically-classified number counts to $I < 22$ and $I < 24$ respectively, using the post-refurbishment HST Medium Deep Survey (MDS) data. With just three categories: spirals, spheroidals and irregulars they found that those objects classified as either spiral or spheroidal had counts consistent with a non-evolving slope, while the irregular sample had much steeper counts and would therefore dominate at fainter magnitudes. If this is the case then additional information about this population and it characteristics at high redshift is necessary to obtain a coherent view of faint galaxy evolution. A categorisation in terms of scale size may be a useful tool to study the evolution of distant field galaxies, especially the growth of spiral disks if such an event is visible. The use of such information to distinguish between various evolutionary models is discussed from a theoretical basis by Im *et al.* (1995). Using pre-refurbishment MDS data Mutz *et al.* (1994) studied the scale sizes of galaxy images as a function of redshift. They find their disk galaxies have intrinsic sizes similar to those seen for spirals locally, although their data is also consistent with size evolution of $(1+z)^{-1}$. Either way there is only weak evidence for size evolution of spiral disks over the last $\sim 3$ Gyrs.

Unfortunately, the Mutz *et al.* (1994) sample only probes to $I \sim 21$, corresponding to a median depth of $z \sim 0.2$ and while the Glazebrook *et al.* (1995b) and Driver *et al.* (1995) are deeper they have only statistical redshift information. Ideally we would like to study the distributions of morphology and size at earlier times to observe the changes in the nature of the faint field population responsible for the steep optical counts. However, such a search is hampered by the faintness of the galaxies which limits redshift identification. Thus at high redshift we are constrained to study only the intrinsically brightest and hence possibly, least representative objects. Gravitationally lensed features, in particular giant arcs may offer an alternative approach to study a small sample of "normal" distant galaxies.

Giant arcs are formed by the serendipitous alignment of a massive, concentrated foreground cluster along our line of sight to a distant background galaxy. The strong deflection of the light paths to the distant galaxy result in our observing a highly distorted and magnified image of the source. The amplification of the source brings what would have been apparently very faint galaxies within the reach of modern spectrographs, enabling us to spectroscopically identify the source redshifts. Moreover, by combining the natural magnification of the source by the lensing cluster with the high resolution imaging available with HST we can gain a detailed insight into the morphologies and sizes of a small sample of very distant, field galaxies. The lensing process induces extreme distortions in the image of the distant galaxy and to accurately reconstruct the complete source morphology thus requires detailed numerical lens inversion for each of the clusters (e.g. Wallington *et al.* 1995). Nevertheless, we believe that sufficient information remains to tackle the question of the broad morphologies and sizes of $z \gtrsim 1$ field galaxies with this technique. We can determine the scale sizes of the sources to search for new populations of compact galaxies (*e.g.* Miralda-Escudé & Fort 1993) and also use the structure in the arcs to look at the clumpiness of the star-formation in these very distant galaxies. In the discussion below we compare the predominantly blue arc sources with samples of the dominant local star-forming field galaxy population: spirals. These local samples are mostly observed in blue optical bands and thus are reasonably matched to our data.

When using arcs to study the sizes of faint galaxies we have to understand both the gross properties of the lensing clusters (see section 4) and also the role of selection biases in the original discovery of the arcs. All of the arcs discussed here were discovered in optical passbands (sampling the blue and ultraviolet restframe) and are effectively surface-brightness selected. It has been demonstrated, however, that these arcs have optical and optical-infrared colours compatible with the bulk of the faint field population (Smail *et al.* 1993) when allowance is made for the lensing amplification. Thus at least from their colours there is no evidence for selection biases peculiar to the lensing process and the arcs appear to represent a random cross section of the high redshift field galaxy population.

If we wish to use the giant arcs to study the scale sizes of distant galaxies we must also discuss the arc detection as a function of intrinsic source size. For the giant arcs used here the high tangential amplification means there should be no bias against finding thick arcs associated with large sources (e.g. Wu & Hammer 1993). However, when dealing with very compact sources we might expect more difficulty with identification of such features because they will appear as separate images of the source, rather than an extended arc. Hence, we would miss such objects unless they contain easily identifiable structures and only then if they are imaged at sufficiently high resolution (e.g. Smail *et al.* 1995a). In conclusion, we would expect to be able to derive a firm upper limit on the scale size distribution of distant galaxies.

This paper presents high resolution imaging of eight spectroscopically confirmed high redshift field galaxies. These galaxies appear as giant arcs, pairs or arclets in the fields of moderate redshift clusters. Using HST we have acquired deep optical imaging of these arcs at 0.1 arcsec resolution. Thus we have detailed views of these galaxies at restframe ultraviolet-blue wavelengths on spatial scales of $\sim$1 kpc.[†] The bulk of the observations of lensed features presented here has not been discussed previously and we show a number of gravitationally lensed objects uncovered with HST which have important consequences for the modelling of the cluster mass distributions.

---

[†]We take $H_{\circ} = 50$ kms/sec/Mpc and $q_{\circ} = 0.5$ unless otherwise stated

## 2 Observations and Reduction

Four of the five clusters presented (A370, AC114, Cl0024+16, A2218) were observed for studies of galaxy evolution in distant clusters (Dressler *et al.* 1994, Couch *et al.* 1994). The presence of giant arcs in these fields is purely serendipitous and in some respects reflects the prevalance of such features in these rich, moderate redshift clusters. The fifth cluster Cl2244−02 was observed in Cycle-2 to study the arc present in the cluster and these images were retrieved from the STScI archive (Prop.# 2801, P.I. Mr. R.E. Sterner). The majority of the clusters discussed here were observed with pre-refurbishment WFC-1 on HST (Table 1). Nevertheless, the relatively long exposure times mean that most of the arcs are sufficiently well detected that we can use the diffraction-limited core of the aberrated HST point spread function (PSF) to obtain 0.1 arcsec resolution images, without resorting to image restoration techniques. Thus HST provides the crucial factor of $\gtrsim$ 4-5 improvement in resolution over the best ground-based imaging (e.g. Kneib *et al.* 1993) necessary for a detailed examination of these distant galaxies.

Table 2 gives half-light radii ($r_{hl}$) for the arcs measured from our HST frames. For the WFC-1 data the images were first processed by deconvolving for 40 iterations of the Lucy-Richardson algorithm using model PSF's created with TINYTIM (Krist 1992) a procedure well-tested by Windhorst *et al.* (1994), who state that it is adequate for restoring the galaxy's light profile. To measure the half-light radii we extract profiles through the arcs orthogonally to the local shear direction indicated by the arc's shape. These effectively one dimensional slices are then corrected to give standard half-light radii with typical measurement errors of $\lesssim$ 0.1 arcsec. It should be noted that for all arcs, except A5 in A370, we detect the source out to 3× the measured half-light radius. To estimate our resolution limit we have analysed similarly processed stars from our images, showing that the resolution limit of HST is approximately $r_{hl} \sim$ 0.1–0.2 arcsec. Thus HST well resolves all but one arc, Cl2244−02, in our sample.

## 3 Results

We discuss lensed features in four of the clusters from our sample (Cl0024+16, A370, A2218, Cl2244−02), where the new HST data contributes significant new information compared to published ground based studies. When the background source has complex structure the improved resolution available from HST graphically confirms the lensed nature of the giant arcs. Some of the HST observations of lensed features used here have been discussed previously, AC114 in Smail *et al.* (1995a), the arclet population in A2218 in Kneib *et al.* (1995b) and the WFC-1 image of Cl0024+16 was commented on by Kassiola *et al.* (1994, 1995).

### 3.1 Cl0024+16

The arcs in Cl0024+16 were originally noted by Koo (1988) and the cluster has been subsequently studied by the Toulouse group. However, their deep spectroscopy has only been able to place limits on the likely source redshift, $1 \lesssim z_{arc} \lesssim 2$ (Mellier *et al.* 1991). We show the three main sections of the arc A/B/C in Figures 1. Unfortunately, all three sections are contaminated by projected bright galaxies. The difficulty of removing this contamination is illustrated by the disagreement between Kassiola *et al.* (1995) and Wallington *et al.* (1995) over which components of the arcs A–C are multiply-imaged using ground-based imaging. The fourth arc, D, has been proposed as a counter arc to A/B/C on a number of occasions, but has usually been dismissed when it can not be easily recreated in the adopted lensing models of the cluster (e.g. Kassiola *et al.* 1992, 1995; Wallington *et al.* 1995).

The factor of $\gtrsim$ 5 improvement in resolution with HST over the ground based data makes the identification of the various multiply imaged components in all the arcs much less ambiguous. We see a bright elongated knot (marked as 2) in all four images A–D, surrounded by a low surface brightness halo. Closer inspection indicates that the knot comprises two peaks, with separations ($A_2$–$D_2$): 1.1, 0.5, 0.7 and 0.6 arcsec, roughly consistent with the relative lengths of the various arc components. In addition to this double peak a fainter feature (marked as 3) is also visible in all four arcs. The parities shown by these features in the individual arcs are consistent with the lensing hypothesis, and indicate that D is a counter-arc to A/B/C, a conclusion which is also supported by the spectroscopy of the four components, as stated in Mellier *et al.* (1991).

A counter-arc such as D arises naturally in models of the triple-arc A–C using a lensing potential with a small core radius and ellipticity. To illustrate this we have devised a simple lensing model for this cluster using one potential for the cluster and perturbations associated with the brighter galaxies near the arc components. The technique and model potentials are the same as those used in Kneib *et al.* (1995a). We find that the arc can be adequately modelled using a nearly-circular mass, $\epsilon \sim 0.15 \pm 0.05$, with a major-axis angle within 10 degrees of the direction to arc D and centred within 2–3 arcsec of the optical centre of the cluster. The cluster has a core radius of $r_c \sim 40 \pm 10$ kpc, this estimate allows us identify a candidate for the fifth image of the background source, object E in Figure 1a. The small core radius for this cluster is comparable to those derived for other well-constrained cluster lenses (e.g. MS2137−23, Mellier *et al.* 1993; A370, this work). For a source at $z_{arc} \sim 1.3$ the mass inside the radius of the arc is $M = 3.0 \pm 0.1\ 10^{14} M_\odot$. Thus not only does the detailed morphology of D support its identification as a counter-arc, but we are also able to easily reproduce its position and morphology using a simple model of the lensing cluster.

### 3.2 A370

The giant arc, A0, in A370 was the first gravitationally lensed galaxy recognised as such (Soucail *et al.* 1987). The arc is shown in Figure 2, its highly elongated morphology implies that the arc consists of multiple images of a single source. With the exceptional seeing of their deep CFHT images Kneib *et al.* (1993) confirmed this hypothesis by identifying the breaks and peaks associated with the multiple images (Figure 2a). However, the situation is complex as not all of the source is multiply-imaged. In the Kneib *et al.* (1993) model only the central P2–P4 are multiple images, with P1 and P5 being singly-imaged regions of the source. Moreover, fine structure is visible within Peak 1, including an apparent bulge/nucleus and faint spiral structure. We thus claim morphological confirmation of the spiral identification for the source.

The extensive studies of the Toulouse group (Fort *et al.* 1988, Mellier *et al.* 1991, Kneib *et al.* 1993) have provided a number of other gravitationally lensed features in this cluster. These include the arclet, A5, as well as a number of candidate multiply-imaged sources (e.g. B2/B3/B4, Figure 2). Among the latter features is the first object (B2/B3/B4) whose redshift, $z_{arc} = 0.865$, has been determined by direct inversion of a lensing model for a cluster (Kneib *et al.* 1993). Unfortunately, these objects are too faint for us to measure reliable scale sizes from our current images and so we will not discuss them further.

One feature visible on our WFC-1 image which has not been previously catalogued is a radial arc (R, Figure 2) – although 4.8 arcsec long, the faintness ($R \sim 22.9$) compared to the cD envelope and the fineness of this feature makes it invisible on even the superb CFHT images of the cluster (Kneib *et al.* 1993). Radial arcs are formed by two images merging across the inner critical curve of the cluster, roughly corresponding to the core radius of the potential. Thus the existence of a radial arc immediately proves that the potential is non-singular. Furthermore, the position of R, lying between 35–60 kpc from the centre of the southern D galaxy, gives a simple estimate of the core size. We estimate therefore a core radius for the sub-clump around the southern D galaxy in A370 of $r_c \sim 50$ kpc. This is similar to the values obtained from lensing analyses of other clusters, which give $r_c \lesssim 50$–100 kpc (c.f. §3.1).

Reassuringly, the lensing model of Kneib *et al.* (1993) which explains the multiply-imaged system B2/B3/B4 is also capable of fitting the radial arc with no modifications. The model predicts that the source is at $z_{arc} \sim 1.3 \pm 0.2$. The model also predicts that this feature corresponds to two merging images from a five-image configuration. The three other images will have a similar surface brightness to the radial arc but smaller angular extent, making them too faint to be detected in the current data. Identification of the other images in deeper HST data will strongly constrain the core radius of the mass distribution in this cluster.

### 3.3 A2218

This cluster was first recognised as a strong gravitational lens by Pelló *et al.* (1988). They subsequently uncovered a large number of candidate arcs and arclets in the cluster core (Pelló *et al.* 1992) and spectroscopically confirmed two of the brighter candidates arcs. We have followed the naming convention used by Pelló *et al.* (1992).

Object LPS# 289 is a typical very blue, high redshift arc (Figure 3a). The arc lies close to a bright subclump within the cluster, in the saddle between the subclump and the dominant cluster galaxy. One end of the arc has high surface brightness and our WFPC-2 image uncovers a wealth of detail in this feature (Figure 3a). The remainder of the arc has much lower surface brightness and also lies in the halo of the subclump. We propose that the arc consists of a singly-imaged bright section (Section 1) and a fainter area of the source which is multiply-imaged (Sections 2 & 3). This configuration is thus very similar to that seen with the arc A0 in A370, although in this case the source morphology appears less regular. The singly-imaged region shows many bright unresolved areas in a complex pattern, all embedded in a lower surface brightness halo, roughly 7.5 arcsec long.

In contrast to LPS# 289, LPS# 359 has a very smooth appearance (Figure 3b). This arc also differs from LPS# 289 in its relatively red colours (Table 2). The main section of the arc is 6.8 arcsec long and the brightness of this arc and its distance from the cluster centre both imply that it is composed of two images of the background source (Kneib *et al.* 1995a,b). If this is the case at least one other image of the source ought to be apparent. We identify two candidates for this third image (labelled A and B in Figure 3b). The optical colours of both of these objects are consistent with those of LPS# 359, as are the peak surface brightnesses and widths (widths $r_{hl} \sim 0.4$ arcsec, lengths $\sim 1.5$ arcsec). The preferred third image within the framework of the Kneib *et al.* (1995a) model is A.

We finally wish to comment on the object LPS# 384, for which inconclusive spectroscopy has been obtained. Pelló *et al.* (1992) present photometry and a low signal to noise spectrum of this candidate arc. They conclude that the source is most likely foreground of the cluster and therefore not lensed, although they also discuss the possibility that the object is a very high redshift gravitationally lensed source, $z_{arc} \gtrsim 2.6$. The HST image of this object leaves no doubt that it is gravitationally lensed (Figure 3c). The arc being formed by two merging images of the background source, the parity of the two images is graphically illustrated by the mirrored structures in the two images. These structures allow us to unambiguously identify a third image (LPS# 468) which we show in Figure 3c. The WFPC-2 image of A2218 provides a large number of similar multiply-imaged sources (e.g. LPS# 289, 323, 384, 730 in Figures 3). The combination of all these multiply-imaged features and the detailed morphological information on each one available from HST makes this cluster one of the best hopes for constraining the redshift distribution of the faint field population from gravitational lensing (Kneib et al. 1995b).

In addition to the arcs in this cluster Pelló et al. (1992) uncovered an unusual ring-like feature around one of the cluster galaxies (#373, Figure 3b). They suggested that this might be an Einstein ring and were able to recreate the feature under this hypothesis, although the contrived lensing geometry led later groups to discount the lensing model (Kassiola & Kovner 1993). Recent spectroscopic studies also tend to weigh against the lensing explanation for this feature (Le Borgne & Pelló reported by Fort & Mellier 1994). Our deep WFPC-2 exposure plainly shows weak spiral structure in the ring, including a dust lane crossing the arc LPS# 359. This gives unequivocal observational evidence against the gravitational lensing explanation of this feature and we conclude that it is a faint red stellar disk associated with LPS# 373.

### 3.4 Cl2244−02

The remarkable circular structure in this cluster was discovered by Lynds & Petrosian (1989) and was spectroscopically confirmed as a gravitationally lensed arc using very long integrations and a novel circular spectroscopic slit (Mellier et al. 1991). The redshift determination comes from identification of a line at the extreme blue limit of the spectrograph, assumed to be Ly$\alpha$ giving $z_{arc} = 2.237$. Irrespective of the reality or identification of this line the continuum shape of the source means it must nevertheless be at relatively high redshift ($z_{arc} \gtrsim 1$). Subsequent imaging observations presented in Hammer et al. (1989) showed distinctive structure within the arc and we mark their features on the WFC-1 image in Figure 4. The six surface brightness peaks along the arc have been explained in the model of Hammer & Rigaut (1989) as arising from two images (P1-P3 and P4-P6) of a single background source, which itself contains 3 peaks. They also identify 2 other possible images of the source: S and T. In this model the two images P1-P3 and P4-P6 ought to have reversed parities and this has been confirmed from optical-infrared colour gradients (Smail et al. 1993). This study also uncovered the first infrared-selected gravitational arc in this cluster, but unfortunately the current exposure is too short to detect this feature.

In the light of the proposed model of this arc one feature in the arc bears comment, the relative brightness of P2 compared to the other structures in the arc. The mirror symmetry postulated for the arc is broken by the relative brightness of P2 compared to its mirror image, P5. The peak surface brightness of P2 is a factor of $\sim 2$ higher than the similar region of P5. Three possible explanations exist for this discrepancy. Firstly, the bright peak may be a superimposed faint star/galaxy, although the similarity of the colours of P2 and P5 mean the additional source must be fairly blue. Secondly, the source seen as P2/P5 may have varied in brightness (e.g. due to a supernova), which coupled to the different light travel times from the observer to source for the two images would lead to one image being brighter than the other for a period of time. The last possibility is the most interesting, this is that a small region of arc around P2 is being amplified by a superimposed massive dark object within the cluster, as discussed by Fort & Mellier (1994). New post-refurbishment observations of this arc are being obtained and they should conclusively determine the reason for P2's brightness.

### 4 Discussion

The aim of this study is to gain a first view of the morphologies and sizes of typical star-forming field galaxies at cosmologically interesting look-back times, back to $z \sim 1$–2 or from 8 to 11 Gyrs ago. In particular we wish to determine: 1) the average sizes of high redshift galaxies, 2) whether all the local galaxy types are present in the high redshift field, 3) whether any new populations exist (e.g. Giraud 1992, Miralda-Escudé & Fort 1993).

It is clear that the arc sources demonstrate a wide variation in morphologies (Table 2, Figures 1–4). The extended sources range from smooth, red systems (e.g. LPS#359 in A2218), through smooth, blue sources (e.g. A5 in A370) to objects with significant internal structure (e.g. Cl0024+16). These characteristics are very similar to what would be observed if we took a sample of the local field population, with the obvious identification of star-forming disk systems and smooth red spheroidals. Within the scheme adopted by Giraud (1992) we find that 50% of our spectroscopic sample show evidence for multiple components within the source. With our current data it is difficult to determine whether this reflects substructure in the lensed galaxy or close interactions and mergers, although the small scales probed ($\lesssim 1$ kpc) would indicate that we are probably seeing individual structures within a single galaxy. Either way it is apparent that the star-formation in the bulk of our sample occurs in distinct regions leading to complex morphologies (e.g. Glazebrook et al. 1995b, Driver et al. 1995).

To study the scale sizes of the sources we must address the possibility that lensing amplification will change the observed source profile. The radial amplification, $A_{rad}$, of a giant arc depends upon the compactness of the lensing potential (Wu & Hammer 1993). For singular isothermal sphere models the amplification is $A_{rad} = 1$, while more compact mass distributions form narrower arcs ($A_{rad} < 1$) and non-singular models thicker arcs ($A_{rad} > 1$). The majority of giant arcs which have been modeled in detail show that the gross properties of the very central regions of the cluster potential can be characterised by a nearly singular isothermal mass distribution with core radii of $\lesssim 50$ kpc (e.g. Kneib et al. 1993, Smail et al. 1995b). In this work we have confirmation of this assumption from modelling of the cluster potential in Cl0024+16, which is supported by the identification of a faint feature in the cluster centre as a fifth image of the source. The presence of such a feature if associated with the giant arc is particularly exciting as this would indicate a compact mass distribution within a cluster which lacks a dominant central galaxy (Miralda-Escudé 1995), although this requires confirmation with deeper HST data.

More direct support comes from the discovery of a candidate radial arc in the well-studied cluster A370. This is only the second such feature discovered and illustrates the advantages of working at high spatial resolution when studying lensed features. The compactness of the lensing potentials derived above shows that although giant arcs are highly elongated in the tangential direction, we expect the radial magnification to be $A_{rad} \sim 1$ within 10–20%. Hence we can adopt the radial light profile of the arc as an unmagnified one dimensional slice through the source.

The half-light radii for our sample are given in Table 2 and illustrated in Figure 5. We have chosen to compare our sample with spirals, the dominant local population of star-forming galaxies. We use, therefore, a local sample taken from Mathewson et al.'s survey (Mathewson et al. 1992) and a more distant sample analysed from WFC-1 MDS images by Mutz et al. (1994). The typical half-light size in the Mathewson et al. sample is $r_{hl} \sim 8.7$ kpc disks, equivalent to a scale length, $r_s = 5.3$ kpc (Mutz et al. 1994). We plot the variation in apparent angular size of such disks as a function of cosmological geometry in Figure 5.

The observed arc widths follow a smooth progression from the scale lengths observed in the lower redshift samples. However, the distribution of arc widths is apparently incompatible with a non-evolving population of disks with intrinsic half-light radii $r_{hl} \sim 8.7$ kpc, irrespective of the adopted geometry. The mean size of the arc sources is $<r_{hl}> = 4.6 \pm 1.5$ kpc. Thus the average arc source is a factor of $\sim 1.5$–2 times smaller than would have been expected from extrapolation of the sizes of local star-forming systems, in standard geometries.

To indicate the strength of evolution required to connect the arc source sizes to local spirals we plot on Figure 5 a model, $r_{hl} \propto (1+z)^{-1}$, which Mutz et al. (1994) state is consistent with their observations, see also Im et al. (1995). A more detailed comparison with the Mutz et al. (1994) sample is hampered by the small samples, but what is apparent, however, is the absence of the larger sources ($r_{hl} \gtrsim 10$ kpc) relative to smaller sources in the arc catalogue. As we discussed earlier this cannot be explained by selection effects in the original sample and we must therefore accept that either: 1) large galaxies with very extended star-forming regions are rarer at $z \sim 1$–2 than they are today or 2) the relative proportions of large and small galaxies has changed between $z \sim 1$ and today. The small number of giant arcs precludes us distinguishing between these alternatives. However, the latter possibility is discussed at length in Driver et al. (1995) and Glazebrook et al. (1995b).

These conclusions are similar to those reached by Smail et al. (1993) who used $K$ imaging of a sample of giant arcs, which considerably overlaps that used here, to measure the rest-frame near-infrared luminosities of the sources. They found that the arc sources had sub-$K^*$ luminosities compared to the local field. One explanation for this difference is a lack of massive, luminous galaxies at $z \gtrsim 1$. Alternatively, this result could arise from a change in the relative abundance of dwarf galaxies compared to more massive systems. Thus the distributions of both $K$ luminosities and sizes of $z \gtrsim 1$ field galaxies indicate that dominance of large, massive galaxies is a relatively recent feature of the field population.

The power of using arcs and arclets for studying the evolution of the sizes of faint galaxies comes from the large redshift range probed. As Kneib et al. (1995b) show it is possible, using HST observations of a well-constrained lensing cluster, to study the redshift distribution of very faint galaxies, $R \lesssim 26$–27, from the shear induced in the galaxy images by the cluster. In principle with deep enough data this technique can be extended to study the distribution of scale sizes as a function of redshift for large samples of very faint galaxies, well beyond the reach of conventional spectroscopy.

# 5 Conclusions

- We have analysed deep pre- and post-refurbishment images of 8 spectroscopically-confirmed giant arcs, arclets and pairs to study the morphologies and sizes of typical galaxies in the $z \gtrsim 1$ field population.

- In addition our high resolution imaging has enabled us to improve the mass models of two of our clusters, Cl0024+16 and A370, from modelling of multiply-imaged background sources and the identification of candidate arcs in the cores of both clusters. Both of our models indicate very compact, $r_c \sim 50$ kpc, mass distributions in these clusters, in agreement with the limits derived by other workers. However, the result for Cl0024+16 is particularly interesting as this cluster lacks a *single* central dominant galaxy.

- We conclude that star-forming galaxies at high redshift are considerably ($\sim 1.5$–$2\times$) more compact than the bulk of the local (spiral) population. If these systems are typical spiral disks then this result argues for evolution in the disk sizes over the last $\sim 10$ Gyrs. A similar conclusion was drawn from near-infrared imaging of the giant arcs which showed that the bulk of the $z \gtrsim 1$ sources are sub-$L^*$ compared to the local field (Smail *et al.* 1993). Both of these results can be explained by either an absence of large galaxies from the $z \sim 1$ field population or a shift in the ratio of massive/dwarf galaxies. Either possibility points towards a considerable change in the nature of the field population since $z \sim 1$.

- At the moment this technique is limited by the small sample of giant arcs available. This obstacle can be removed by using very deep HST imaging of a small number of well-constrained lensing clusters. The large samples of arclets produced can be analysed to provide both the redshift and scale size information needed to conclusively tackle the question of the size evolution of typical galaxies at high redshift.

## Acknowledgements

We thank Roger Blandford, Bernard Fort, David Hogg, Yannick Mellier and Jordi Miralda-Escudé for useful discussions. We also wish to thank Dr. S. Mutz for providing the data from his HST study. Support via a NATO Advanced Fellowship and a Carnegie Fellowship (IRS) is gratefully acknowledged. WJC acknowledges support from the Australian Department of Industry, Science and Technology, the Australian Research Council and Sun Microsystems.

# 5 References

Couch, W.J., Ellis, R.S., Sharples, R.M. & Smail, I., 1994, ApJ, 430, 121.

Dressler, A., Oemler, A., Butcher, H. & Gunn, J.E., 1994, ApJ, 430, 107.

Driver, S.P., Windhorst, R.A., Ostrander, E.J., Keel, W.C., Griffiths, R.E. & Ratnatunga, K.U., 1995, ApJL, submitted.

Ellis, R.S., 1994, in *Stellar Populations*, eds van der Kruit, P. & Gilmore, G., IAU Symposium 164.

Fort, B., Prieur, J.-L., Mathez, G., Mellier, Y. & Soucail, G., 1988, A&A, 200, 17.

Fort, B. & Mellier, Y., 1994, A&A Rev., 5, 239.

Giraud, E., 1992, A&A, 257, 501.

Glazebrook, K., Ellis, R.S., Colless, M.M., Broadhurst, T.J., Allington-Smith, J. & Tanvir, N.R., 1995a, MNRAS, in press.

Glazebrook, K., Ellis, R.S., Santiago, B. & Griffiths, R.E., 1995b, Nature, submitted.

Hammer, F., Le Fevre, O., Jones, J., Rigaut, F. & Soucail, G., 1989, A&A, 208, L7.

Hammer, F. & Rigaut, F., 1989, A&A, 226, 45.

Im, M., Casertano, S., Griffiths, R.E. & Ratnatunga, K.U., 1995, ApJ, in press.

Kassiola, A., Kovner, I. & Fort, B., 1992, ApJ, 400, 41.

Kassiola, A. & Kovner, I., 1993, ApJ, 417, 474.

Kassiola, A., Kovner, I., Fort, B. & Mellier, Y., 1994, ApJL, 429, L9.

Kassiola, A., Kovner, I. & Dantel-Fort, M., 1995, MNRAS, submitted.

Kneib, J.-P., Mellier, Y., Fort, B. & Mathez, G., 1993, A&A, 273, 367.

Kneib, J.-P., Mathez, G., Fort, B., Mellier, Y., Soucail, G. 1994, A&A, 286, 701.

Kneib, J.-P., Mellier, Y., Melló, R., Miralda-Escudé, J., Le Borgne, J.-F., Böhringer, H. & Picat, J.-P., 1995a, A&A, in press.

Kneib, J.-P., Ellis, R.S., Smail, I., Couch, W.J. & Sharples, R.M., 1995b, ApJL, submitted.

Koo, D.C., 1988, in *Large-Scale Motions in the Universe*, eds, Rubin, V.G. & Cayne, G.V., Princeton Univ. Press, p513.

Krist, J., 1992, TINYTIM User's Manual, STScI, Baltimore.

Lilly, S.J., 1993, ApJ, 411, 501.

Lynds, R. & Petrosian, V., 1989, ApJ, 336, 1.

Mathewson, D.S., Ford, V.L. & Buchhorn, M., 1992, ApJS, 81, 413.

Mellier, Y., Fort, B., Soucail, G., Mathez, G. & Cailloux, M., 1991, ApJ, 380, 334.

Mellier, Y., Fort, B. & Kneib, J.-P., 1993, ApJ, 407, 33.

Miralda-Escudé, J. & Fort, B., 1993, ApJ, 417, L5.

Miralda-Escudé, J., 1995, 438, 514.

Mutz, S.B., Windhorst, R.A., Schmidtke, P.C., Pascarelle, S.M., Griffiths, R.E., Ratnatunga, K.U., Casertano, S., Im, M., Ellis, R.S., Glazebrook, K., Green, R.F. & Sarajedini, V.L., 1994, ApJL, 434, L55.

Pelló, R., Le Borgne, J.F., Sanahuja, B., Mathez, G. & Fort, B., 1988, A&A, 190, L11.

Pelló, R., Le Borgne, J.F., Sanahuja, B., Mathez, G. & Fort, B., 1992, A&A, 266, 6.

Smail, I., Ellis, R.S., Aragòn-Salamanca, A., Soucail, G., Mellier, Y. & Giraud, E., 1993, MNRAS, 263, 628.

Smail, I., Ellis, R.S. & Fitchett, M.J., 1994, MNRAS, 270, 245.

Smail, I., Couch, W.J., Ellis, R.S. & Sharples, R.M., 1995a, ApJ, 440, 501.

Smail, I., Hogg, D.W., Blandford, R.D., Cohen, J.G., Edge, A.C. & Djorgovski, S.G., 1995b, MNRAS, submitted.

Soucail, G., Fort, B., Mellier, Y. & Picat, J.-P., 1987, A&A 172, 14.

Tresse, L., Hammer, F., LeFevre, O. & Proust, D., 1993, A&A, 277, 53.

Wallington, S., Kochanek, C.S. & Koo, D.C., 1995, ApJ, submitted.

Windhorst, R.A., Schmidtke, P.C., Pascarelle, S.M., Gordon, J.M., Griffiths, R.E., Ratnatunga, K.U., Neuschaefer, L.W., Ellis, R.S., Gilmore, G., *et al.*, 1994, AJ, 107, 930.

Wu, X.P. & Hammer, F., 1993, MNRAS, 262, 187.

**Figures**

**Figure 1:** (a) The central region of our WFC-1 F702W image of Cl0024+16 ($z_{cl} = 0.39$). The compact core of the cluster is easily visible as are the four components of the arc. We overlay on this image the critical curves and the shear map from our lensing model of the cluster, see §3.1. Using this model we tentatively identify object E (Kassiola *et al.* 1992) as a possible fifth image of the source. (b) An enlarged view of the four components of the arc, A–D. The morphological features which repeat between all the arc segments are shown, conclusively proving that component D is the counter-arc to A–C. The images are 15×15 arcsec and have been smoothed with a gaussian with a 0.15 arcsec FWHM to reduce sky-noise. Marked are the features discussed in the text using the naming scheme adopted by Kassiola and co-workers (Kassiola *et al.* 1992, 1995).

**Figure 2:** The central section of the WFC-1 F702W image of A370 ($z_{cl} = 0.37$) showing the giant arc A0. We identify the two singly-imaged regions P1 and P5 as well as the multiply-imaged sections P2–P4. Notice the bulge and weak spiral structure visible in P1. Also shown is the newly discovered radial arc, R, for which we estimate a redshift $z_{arc} = 1.3 \pm 0.2$ and the multiply-imaged feature B2/B3/B4 modeled by Kneib *et al.* (1993).

**Figure 3:** (a) The arc LPS# 289 in A2218 ($z_{cl} = 0.18$). The arc has a spectroscopic redshift of $z_{arc} = 1.038$ and a large amount of structure is visible in the main component (section 1). Sections 2 and 3 may be multiply-imaged regions of the source. Also shown is the multiply-imaged arc LPS# 730, for which a redshift has not yet been measured. (b) The figure shows the Einstein ring proposed by Pelló *et al.* (1992) around galaxy LPS# 373 in A2218. Weak spiral structure is visible within the ring, supporting our conclusion that it is associated with the central galaxy rather than a lensed feature. The red arc LPS# 359 shows little internal structure, consistent with an identification as a moderate redshift spheroidal system. (c) The left panel shows the candidate arc LPS# 384 which Pelló *et al.* (1992) claimed to most likely be foreground of the cluster from weak spectroscopic evidence. The multiple nature of this arc is striking and we thus claim that it arises from background source lensed by the cluster core. The right panel has a transposed image of the counter arc (LPS# 468) to LPS# 384 showing similar internal structure to the arc.

**Figure 4:** The core of the compact cluster Cl2244−02 at $z_{cl} = 0.33$. The partial Einstein ring is clearly visible, as are various surface brightness peaks along this arc. We use the naming scheme of Hammer *et al.* (1989) to label these features. The objects S and T were proposed by Hammer *et al.* (1989) as possible additional images of the background source at $z_{arc} = 2.238$.

**Figure 5:** The distribution of observed half-light radii versus redshift for the arc sample (•). We assume no radial magnification by the lensing process. We plot for comparison the half-light radii of spiral galaxies from both local (Mathewson *et al.* 1992) and moderate redshift samples (□, Mutz *et al.* 1994). Finally, we show the observed size of a typical local spiral galaxy, $r_{hl} = 8.7$ kpc, as a function of redshift in different cosmologies and the effect of a scale evolution of the form $r_{hl} \propto (1+z)^{-1}$ in a $q_o = 0.5$ cosmology. The relative dearth of large sources over smaller systems, compared to that expected from a non-evolving population of sources is readily apparent. This plot is adapted from Mutz *et al.* (1994).

**Tables**

**Table 1:** Observational details of the WFC-1 and WFPC-2 imaging of giant arcs.

**Table 2:** Basic photometric data on the spectroscopically confirmed lensed features in the HST fields. The data are taken from the compilations of Mellier *et al.* (1991), in addition to Kneib *et al.* (1993, 1995a), Belló *et al.* (1992) and Smail *et al.* (1993). The last two columns give the half-light widths of the arc sources, $r_{hl}$, in arcseconds and whether multiple components are visible in the light distribution.